\documentclass[dvips]{article}
\usepackage{graphicx}
\usepackage{amssymb}
\usepackage{graphicx}
\usepackage{dcolumn}
\usepackage{amsmath}
\usepackage{lscape}
\usepackage{longtable}
\usepackage{subfigure}
\usepackage{color}

\begin{document}
\newcommand{\bd}{\begin{document}}
\newcommand{\ed}{\end{document}}
\newcommand{\bc}{\begin{center}}
\newcommand{\ec}{\end{center}}
\newcommand{\bfr}{\begin{flushright}}
\newcommand{\efr}{\end{flushright}}
\newcommand{\lt}{\left}
\newcommand{\rt}{\right}
\newcommand{\vs}{\vspace}
\newcommand{\hs}{\hspace}
\newcommand{\beq}{\begin{equation}}
\newcommand{\eeq}{\end{equation}}
\newcommand{\lb}{\linebreak}
\newcommand{\pb}{\pagebreak}
\newcommand{\mb}{\makebox}
\newcommand{\fb}{\framebox}
\newcommand{\mc}{\multicolumn}
\newcommand{\ben}{\begin{enumerate}}
\newcommand{\een}{\end{enumerate}}
\newcommand{\bit}{\begin{itemize}}
\newcommand{\eit}{\end{itemize}}
\newcommand{\ol}{\overline}
\newcommand{\un}{\underline}
\newcommand{\lefq}{\lefteqn}
\newcommand{\ba}{\begin{array}}
\newcommand{\ea}{\end{array}}
\newcommand{\beqa}{\begin{eqnarray}}
\newcommand{\eeqa}{\end{eqnarray}}
\newcommand{\beqas}{\begin{eqnarray*}}
\newcommand{\eeqas}{\end{eqnarray*}}
\newcommand{\bfg}{\begin{figure}}
\newcommand{\efg}{\end{figure}}
\newcommand{\bds}{\begin{displaymath}}
\newcommand{\eds}{\end{displaymath}}
\newcommand{\btb}{\begin{tabbing}}
\newcommand{\etb}{\end{tabbing}}
\bc {
\textbf{\huge Non-Hermitian Dirac Hamiltonian in three dimensional gravity and pseudo-supersymmetry} } \ec

\vs{1cm}

\bc
{\it \"Ozlem Ye\c{s}ilta\c{s}$^{*}${\footnote {e-mail : yesiltas@gazi.edu.tr}   \\
$^{*}$Department of Physics, Faculty of Science, Gazi University,
06500 Ankara, Turkey\\
\vspace{.16cm}

}} \ec \vs{1cm}
\begin{abstract}
The Dirac Hamiltonian in the $2+1$ dimensional curved space-time has been studied with a metric for an expanding de Sitter space-time which is a two sphere. The spectrum and the exact solutions of the time dependent non-Hermitian and angle dependent Hamiltonians are obtained in terms of the Jacobi and Romanovski polynomials. Hermitian equivalent of the Hamiltonian  obtained from the Dirac equation is discussed in the frame of pseudo-Hermiticity. Furthermore, pseudo-supersymmetric quantum mechanical techniques are expanded to a  curved Dirac Hamiltonian and a partner curved Dirac Hamiltonian is generated. Using $\eta$-pseudo-Hermiticity, the intertwining operator connecting the non-Hermitian Hamiltonians to the Hermitian counterparts is found. We have obtained a new metric tensor related to the new Hamiltonian.
\end{abstract}
\noindent {\bf keyword:}   Dirac equation, curved space-time, non-Hermitian Hamiltonians \\


\section{Introduction}
Two great achievements of the twentieth century, quantum mechanics and general relativity  are very successful in their own  boundaries to describe the nature, however they are  incompatible, for instance they break down at extremely tiny distance which is  Planck scale. In modern physics, the unified theory of gravitation and quantum mechanics which plays a fundamental role with exactly solvable gravitational field equations  has always attracted considerable interest. On the other hand, experimental studies have been performed for the gravitational effects in quantum theory; earth's rotational effect on the phase of the neutron wave function \cite{11}, experimental nano-diamond interferometry \cite{22}, quantum light in coupled interferometers \cite{33}.In theoretical physics, interesting works including mathematical aspects of the Dirac Hamiltonians with their spectrum has been investigated such as shifted energy levels of the hydrogen atom in a region of curved space-time \cite{parker},  modified super-symmetric harmonic oscillator on a two dimensional gravitational field \cite{moayedi}, exact solutions in a $2+1$ dimensional contracting and expanding curved space-time \cite{sucu}, singularities in $2+1$ dimensional spacetime \cite{halilsoy}, Hawking radiation of particles from a black hole \cite{BH}. Curved Dirac systems have also dynamical symmetries \cite{vercin}, considering the $(2+1)$ dimensions $SL(2,c)$ symmetries of the  Dirac Hamiltonian  have been studied \cite{moayedi}. In non-relativistic domain, $\mathcal{PT}$ symmetric theories which are examining the non-Hermitian Hamiltonians with complex potentials and real eigenvalues have been attracted much interest \cite{bender}. Here $\mathcal{P}$ and $\mathcal{T}$ are the parity and time-reversal operators whose action on position wave functions $\psi(x)$  can be shown by $\mathcal{P}\psi(x)=\psi(-x)$ and $\mathcal{T}\psi(x)=\psi^{*}(x)$. More general concept of the Hermiticity is found as a generalization of the $\mathcal{PT}$ symmetry which is called as pseudo-Hermiticity \cite{mos}. The real eigenvalues and corresponding eigen-states of a non-Hermitian Hamiltonian associated with a symmetry such as the $\eta$-pseudo-Hermitian Hamiltonian. It is shown that any inner product may be defined in terms of a metric operator $\eta$. Moreover, hydrogen atom freely falling in a curved spacetime with the curvature effects on the spectrum is investigated \cite{1} and a general scalar product called as Parker product is defined for the Dirac equation in a curved background. It is also shown that if the time dependence of the metric is not omitted, there occur a violation of Hermiticity in the Dirac Hamiltonian \cite{1}. Then, in the light of a weight operator in Parker scalar product, a non-Hermitian Hamiltonian and its Hermiticity without omitting the time dependency of the metric is discussed in \cite{parker1}. The  problem of non-uniqueness and Hermiticity of Hamiltonians using the frame of pseudo-Hermitian Hamiltonian through the stationary gravitational fields and self-conjugacy of Dirac Hamiltonians with some examples are examined in \cite{2}, \cite{3}.

On the further side of these discussions, generalization of the supersymmetric quantum mechanics, i.e. pseudo-supersymmetric quantum mechanics and its effects in curved spacetime hasn't been involved in the literature to our knowledge. Intertwining operators linking non-Hermitian Dirac Hamiltonians and Hermitian counterparts may give rise to a more general Dirac Hamiltonian chains. Accordingly we have examined the Dirac equation in $2+ 1 $ dimensional universe with an  induced  metric in this paper. We have given the real spectrum  of the non-Hermitian Dirac Hamiltonian and corresponding solutions of time dependent and angular parts. Moreover we have shown that a new non-Hermitian Dirac Hamiltonian in curved space-time  can be generated using the aspects of pseudo-supersymmetric quantum mechanics and this may lead to another metric tensor which may be related to the generated Hamiltonian. This paper is organized as follows; Section 2 is devoted to the Dirac equation in $(2+1)$ curved space-time and separation of variables, Section 3 involves the exact solutions and pseudo-supersymmetric quantum mechanical applications are discussed in Section 4. Finally, we conclude the paper in Section 5.

\section{Dirac equation}
The generally covariant form of the Dirac equation is
\begin{equation}\label{1}
    i\gamma^{\mu}(x)(\partial_{\mu}+i e A_{\mu}-\Gamma_{\mu}(x))\Psi(x)=M\Psi(x)
\end{equation}
where $M$ is the mass and $e$ is the charge of the particle, $A_\mu$ is the electromagnetic vector potential, $\Gamma_{\mu}(x)$ is the spin connection,  $\gamma^{\mu}(x)$ are the space-time dependent matrices. The spin connection relation is defined by
\begin{equation}\label{6}
    \Gamma_{\mu}(x)=\frac{1}{4}g_{\lambda \rho}(e_{\nu,\mu}^{a}e_{a}^{\rho}-\Gamma_{\nu \mu}^{\rho})S^{\lambda \nu}
\end{equation}
where $\Gamma_{\nu \mu}^{\rho}$ is the Christoffel symbol. In \cite{c}, the induced metric which is the static form of the Euclidean de Sitter space-time is given by
\begin{equation}\label{2}
    ds^{2}=\ell^{2}d\tau^{2}-\ell^{2}\sinh^{2}\tau d\theta^{2}-\ell^{2}\sinh^{2}\tau\sin^{2}\theta d\phi^{2}
\end{equation}
where $\tau \in [0, \infty)$, $\phi \in [0, 2 \pi)$ and $\theta \in [0,\pi)$, $\ell$ is the radius of the universe. The vierbein matrix has become
\begin{equation}\label{3}
    e^{\mu}_{a}(x)=\left(
                     \begin{array}{ccc}
                       \frac{1}{\ell} & 0 & 0 \\
                       0 & \frac{1}{\ell \sinh\tau} & 0 \\
                       0 & 0 & \frac{1}{\ell\sinh\tau\sin\theta} \\
                     \end{array}
                   \right),
\end{equation}
here $\mu$ labels the general space-time coordinate and $a$ labels the local Lorentz space-time. The vierbein field as the square root of the metric tensor is written as
\begin{equation}\label{4}
    g^{\mu \nu}=e^{\mu}_{a}e^{\nu}_{b}\eta^{ab},
\end{equation}
and
\begin{equation}\label{5}
    \gamma^{\mu}(x)=e^{\mu}_{a}(x) \gamma^{a}
\end{equation}
where $\gamma^{a}$ are constant matrices. Additionally, one can write
\begin{equation}\label{7}
     S^{\lambda \nu}=\frac{1}{2}[\gamma^{\lambda}(x), \gamma^{\nu}(x)].
\end{equation}
The fermions have only one spin polarization in $(2+1)$ dimensions, then the Dirac matrices can be expressed in
terms of the Pauli spin matrices $\gamma^{i}=(\sigma^{3}, i\sigma^{1}, i \sigma^{2})$ and they satisfy the anti-commutation relation which is
\begin{equation}\label{anti}
    \{\sigma^{i}, \sigma^{j}\}=2\eta^{ij}I_{2x2}
\end{equation}
where $\eta^{ij}$ is the $(2+1)$ dimensional Minkowski space-time metric and $I_{2x2}$ is the identity matrix. These matrices can be chosen as $\gamma^{0^\dag}=-\gamma^{0}, \gamma^{i^\dag}=\gamma^{i}$. If we use (\ref{3})- (\ref{anti}), we arrive at
\begin{eqnarray}\label{77}
  \Gamma_{0} &=& 0 \\
  \Gamma_{1} &=& -\frac{1}{2}\cosh\tau \gamma^{0}\gamma^{1} \\
  \Gamma_{2} &=& -\frac{1}{2}(\cosh\tau \sin\theta \gamma^{0}\gamma^{2}+\cos\theta \gamma^{1}\gamma^{2} ).
\end{eqnarray}
Then, (\ref{1}) becomes
\begin{equation}\label{8}
    \left(\gamma^{0}(\partial_{\tau}+\coth\tau)+iM\ell \hat{E}+\frac{\gamma^{1}}{\sinh\tau}(\partial_{\theta}+\frac{1}{2}\cot\theta)+\frac{\gamma^{2}}{\sinh\tau \sin\theta}\partial_{\phi}+ie\ell A_{1}(\tau,\theta)\gamma^{1}\right)\Psi=0.
\end{equation}
It is noted that $\Psi$ depends on $(\tau, \theta, \phi)$ with two components, $\Psi=\left(
                                                                                     \begin{array}{c}
                                                                                       \psi_{1} \\
                                                                                       \psi_2 \\
                                                                                     \end{array}
                                                                                   \right).$ Thus, (\ref{8}) turns into
\begin{equation}\label{9}
 \left(   \frac{\partial}{\partial\tau}+\coth\tau+iM\ell\right)\psi_1+\frac{i}{\sinh\tau}\left(\frac{\partial}{\partial\theta}+\frac{\cot\theta}{2}-
 \frac{i}{\sin\theta}\frac{\partial}{\partial \phi}+ie  A_{1}(\tau,\theta)\right)\psi_2=0,
\end{equation}
\begin{equation}\label{10}
- \left(   \frac{\partial}{\partial\tau}+\coth\tau-iM\ell\right)\psi_2+\frac{i}{\sinh\tau}\left(\frac{\partial}{\partial\theta}+\frac{\cot\theta}{2}+
 \frac{i}{\sin\theta}\frac{\partial}{\partial \phi}+ie  A_{1}(\tau,\theta)\right)\psi_1=0.
\end{equation}
Applying the separation of variables process leads to
\begin{equation}\label{11}
    \left(   \frac{\partial}{\partial\tau}+\coth\tau+iM\ell\right)T_1(\tau)-\frac{i\omega_{2}}{\sinh\tau}T_2(\tau)=0,
\end{equation}
\begin{equation}\label{12}
   - \left(\frac{\partial}{\partial\tau}+\coth\tau-iM\ell\right)T_2(\tau)+\frac{i\omega_{1}}{\sinh\tau}T_1(\tau)=0,
\end{equation}
and
\begin{equation}\label{13}
    \left(-\frac{\partial}{\partial\theta}-\frac{\cot\theta}{2}+\frac{i}{\sin\theta}\frac{\partial}{\partial\phi}-
    ie  A_{1}\right)Y_2(\theta,\phi)=\omega_2 Y_1(\theta,\phi),
\end{equation}
\begin{equation}\label{14}
   \left(\frac{\partial}{\partial\theta}+\frac{\cot\theta}{2}+\frac{i}{\sin\theta}\frac{\partial}{\partial\phi}+
    ie A_{1}\right)Y_1(\theta,\phi)=\omega_1 Y_2(\theta,\phi) .
\end{equation}
Here, $\psi_1=T_1(\tau)Y_1(\theta,\phi)$ and $\psi_2=T_2(\tau)Y_2(\theta,\phi)$ and $A_{\theta}(\tau,\theta)$ is chosen as $A_{\theta}(\tau,\theta)=\sinh\tau A_{1}(\theta)$, $\omega_{1,2}$ are the separation constants. The angular part is defined as
\begin{equation}\label{15}
\left(
                                                                                                                    \begin{array}{c}
                                                                                                                      Y_1(\theta,\phi) \\
                                                                                                                      Y_2(\theta,\phi) \\
                                                                                                                    \end{array}
                                                                                                                  \right)=
                                                                                                                  e^{i m \phi}\left(
                                                                                                                                \begin{array}{c}
                                                                                                                                  \Theta_1(\theta) \\
                                                                                                                                  \Theta_2(\theta) \\
                                                                                                                                \end{array}
                                                                                                                              \right),
\end{equation}
then we have
\begin{equation}\label{16}
    \left(-\frac{\partial}{\partial\theta}-\frac{\cot\theta}{2}-\frac{m}{\sin\theta}-
    ie   A_{1}\right)\Theta_2=\omega \Theta_1
\end{equation}
and
\begin{equation}\label{17}
    \left(\frac{\partial}{\partial\theta}+\frac{\cot\theta}{2}-\frac{m}{\sin\theta}+
    ie   A_{1}\right)\Theta_1=\omega \Theta_2.
\end{equation}
We have used $\omega_1=\omega_2=\omega$. The first order angular and time dependent equations give us
\begin{equation}\label{18}
  -\frac{d^{2}\Theta_{1}(\theta)}{d\theta^{2}}+(-2ie  A_{1}-\cot \theta)\frac{d\Theta_{1}(\theta)}{d\theta}+\left((e   A_{1}-\frac{i\cot \theta}{2})^{2}-m\cot\theta\csc\theta+(m^{2}+\frac{1}{2})\csc^{2}\theta-ie  \frac{\partial A_{1}}{\partial\theta}\right)\Theta_{1}(\theta)(\theta)=\omega^{2}\Theta_{1}(\theta)
\end{equation}
and
\begin{equation}\label{19}
  -\frac{d^{2}\Theta_{2}(\theta)}{d\theta^{2}}+(-2ie  A_{\theta}-\cot \theta)\frac{d\Theta_{2}(\theta)}{d\theta}+\left((e   A_{\theta}-\frac{i\cot \theta}{2})^{2}+m\cot\theta\csc\theta+(m^{2}+\frac{1}{2})\csc^{2}\theta-ie  \frac{\partial A_{1}}{\partial\theta}\right)\Theta_{2}(\theta)(\theta)=\omega^{2}\Theta_{2}(\theta).
\end{equation}
Hence, the time dependent equations are obtained as
\begin{equation}\label{20}
   \frac{d^{2}T_{1}(\tau)}{d\tau^{2}}+3\coth\tau \frac{dT_{1}(\tau)}{d\tau}+(i\ell M\coth\tau+(\omega^{2}-1)\csc h \tau^{2}+\ell^{2}M^{2}+2\coth\tau^{2})T_{1}(\tau)=0,
\end{equation}
and
\begin{equation}\label{21}
   \frac{d^{2}T_{2}(\tau)}{d\tau^{2}}+3\coth\tau \frac{dT_{2}(\tau)}{d\tau}+(-i\ell M\coth\tau+(\omega^{2}-1)\csc h \tau^{2}+\ell^{2}M^{2}+2\coth\tau^{2})T_{2}(\tau)=0.
\end{equation}

\section{Exact Solutions}
Let us see  the bound states and corresponding solutions of the Dirac Hamiltonian.
\subsection{Solutions of angular part}
In order to obtain a Schr\"{o}dinger-like equation, we choose $A_{1}(\theta)=i\frac{\cot\theta}{2e}$ and use in (\ref{18}) and (\ref{19}), then we get
\begin{eqnarray}\label{2si1}
  V_+(\theta) &=& -m \cot \theta \csc \theta+m^{2}\csc^{2}\theta \\
  V_-(\theta) &=& m \cot \theta \csc \theta+m^{2}\csc^{2}\theta \label{2si2}
\end{eqnarray}
where $V_1$ and $V_2$ are the functions of the Hamiltonians which are
\begin{eqnarray}
  h_+\Theta_1 &=& \omega^{2}\Theta_1 \\
  h_-\Theta_2 &=& \omega^{2} \Theta_2
\end{eqnarray}
and
\begin{equation}\label{22}
  h_{\pm}=-\frac{d^{2}}{d\theta^{2}}+V_{\pm}(\theta).
\end{equation}
The partner Hamiltonians can be factorized as
\begin{eqnarray}\label{23}
  h_{-} &=& \textbf{A}^{\dag}\textbf{A},~~~~ h_{+} =\textbf{A} \textbf{A}^{\dag}\\
  \textbf{A} &=& \frac{d}{d\theta}+W(\theta),~~~~\textbf{A}^{\dag}= - \frac{d}{d\theta}+W(\theta). \label{24}
\end{eqnarray}
In our case super-potential $W(\theta)=B \csc\theta-A \cot\theta$, $A, B$ are constants and $V_{\pm}(\theta)=W(\theta)^{2}\mp \partial_{\theta}W$. It is known that (\ref{2si1}) and (\ref{2si2}) are shape invariant potentials when $m\rightarrow -m$ is used in (\ref{2si1}), (\ref{2si2}) can be obtained. If the supersymmetry is unbroken, the ground-state  of $h_{+}$ has zero energy $\omega_{+,0}=0$ . The energy eigenvalues of the partner Hamiltonians are connected by operators in  (\ref{24}). Thus we have,
\begin{equation}\label{25}
  \omega_{+,n}=\omega_{-,n-1},~~n=0,1,...
\end{equation}
The eigenfunctions are related by the operators as
\begin{equation}\label{b}
    \Theta_{-,n-1}=\frac{1}{\sqrt{\omega_{+,n}}} \textbf{A} \Theta_{+,n}.
\end{equation}
In \cite{6}, the potential which is type $(PI)$  given below
\begin{equation}\label{26}
  V_{+}(\theta)=-A^{2}+(A^{2}+B^{2}-A)\csc\theta^{2}-B(2A-1)\csc\theta\cot\theta.
\end{equation}
The solutions are given in \cite{6} for (\ref{26}) which are
\begin{eqnarray}\label{sol1}
  \omega_{+,n}=\omega_{-, n-1} &=&\pm \sqrt{(A+n)^{2}-A^{2}} \\
  \Theta_{+,n}=\Theta_{-,n} &=& N (1-\cos\theta)^{\frac{A-B}{2}}(1+\cos\theta)^{\frac{A+B}{2}}P^{(A-B-1/2; A+B+1/2)}_{n}(\cos\theta)\label{sol2}
\end{eqnarray}
where $P^{(b,c)}_{a}(\emph{x})$ stand for the Jacobi polynomials. For our case,
\begin{equation}\label{27}
  A=\frac{1+2m}{2},~~B=\frac{1}{2}.
\end{equation}
The  normalization constant $N$ is given by
\begin{equation}\label{N}
    N=\frac{2^{m+2}}{2n+m+2}\frac{\Gamma(n+m+3/2)\Gamma(n+3/2)}{n! \Gamma(n+m+2)}.
\end{equation}
And the angular part solutions are
\begin{equation}
\left(
  \begin{array}{c}
    Y_1(\theta,\phi) \\
    Y_2(\theta,\phi) \\
  \end{array}
\right)=e^{im\phi}\left(
          \begin{array}{c}
            \Theta_{-,n-1} \\
            \Theta_{+,n} \\
          \end{array}
        \right).
\end{equation}
\subsection{Solutions of the time dependent part}
Using a mapping that is
\begin{equation}\label{29}
  T_{1,2}(\tau)=\csc h \tau^{\frac{3}{2}} ~y_{1,2}(\tau)
\end{equation}
in (\ref{20}) and (\ref{21}), we get
\begin{equation}\label{30}
 -\frac{d^{2}y_{1}}{d\tau^{2}}+\left(\frac{1}{4}-\ell^{2}M^{2}-i \ell M \coth\tau-(\omega^{2}+\frac{1}{4})\csc h \tau ^{2}\right)y_{1}(\tau)=0
\end{equation}
\begin{equation}\label{31}
 - \frac{d^{2}y_{2}}{d\tau^{2}}+\left(\frac{1}{4}-\ell^{2}M^{2}+i \ell M \coth\tau-(\omega^{2}+\frac{1}{4})\csc h \tau ^{2}\right)y_{2}(\tau)=0.
\end{equation}
Now we have $z=i\coth (i\tau)$ ($-\infty<\tau<\infty$) in above equations, we get
\begin{equation}\label{32}
    -(1+z^{2})\frac{d^{2}y_{k}}{dz^{2}}-2z\frac{dy_{k}}{dz}+(-\frac{\ell^{2}M^{2}-1/4}{1+z^{2}}+\omega^{2}+\frac{1}{4}+\frac{\ell M \epsilon z}{1+z^{2}})y_{k}(z)=0
\end{equation}
where $-\infty < z <\infty$ and we use a more compact form for equations (\ref{30}) and (\ref{31}) and $\epsilon=\left\{
                                                                                         \begin{array}{ll}
                                                                                           -1, & \hbox{$k=1$;} \\
                                                                                           +1, & \hbox{$k=2$.}
                                                                                         \end{array}
                                                                                       \right.$.
Then,  if we put $y_{k}(z)=(1+z^{2})^{-\frac{1}{4}} \bar{y}_{k}(z)$ into (\ref{32}), we obtain
\begin{equation}\label{33}
    -(1+z^{2})\frac{d^{2}\bar{y}_{k}}{dz^{2}}-z\frac{d\bar{y}_{k}}{dz}+(-\frac{\ell^{2}M^{2}}{1+z^{2}}+\omega^{2}+\frac{\ell M \epsilon z}{1+z^{2}})\bar{y}_{k}(z)=0.
\end{equation}
If we give a polynomial solution which is given below
\begin{equation}\label{34}
  \bar{y}_{k}(z)=(z+i)^{-\frac{1}{2}(\emph{A}+i\emph{B})}(z-i)^{-\frac{1}{2}(\emph{A}-i\emph{B})}P(z)
\end{equation}
where $P(z)$ is the unknown polynomial and substituting (\ref{34}) into (\ref{33}), we get
\begin{equation}\label{36}
  (1+z^{2})P^{''}(z)+(z(1-2A)-2B)P^{'}(z)+(A^{2}-\omega^{2})P(z)=0.
\end{equation}
Here $A, B$ are constants and $P(z)$ is the so-called Romanovski polynomials \cite{kirk}, \cite{ques}, \cite{nass} which are solutions of the differential equation given by
\begin{equation}\label{37}
 (1+x^{2}) \frac{d^{2}R(x)}{dx^{2}}+(2bx+a)\frac{dR(x)}{dx}-\bar{\nu}(\bar{\nu}-1+2b)R(x)=0,~~~-\infty< x<\infty,
\end{equation}
where $\bar{\nu}$ is a quantum number $\bar{\nu}=0,1,...$ and $R(x)=R^{(a,b)}_{\bar{\nu}}(x)$ . Here, we note that the constants $A, B$ are given by
\begin{eqnarray}\label{38}
  A &=&\frac{1}{16\ell M \epsilon} (-8\ell M \epsilon+4\sqrt{2}\ell^{2}M^{2}\sqrt{\textbf{a}_{1}-1-4\ell^{2}M^{2}}+\sqrt{2}(1+\textbf{a}_{1})\sqrt{\textbf{a}_{1}-1-4\ell^{2}M^{2}})  \\
  B &=& \sqrt{\frac{\textbf{a}_{1}-1}{8}-\frac{1}{2}\ell^{2}M^{2}}\\
  \textbf{a}_{1}&=&\sqrt{(1+4\ell^{2}M^{2})^{2}+16\ell^{2}M^{2}}\\
  (a,b)       &=& (-2B, \frac{1}{2}-A).
\end{eqnarray}
Then we have
\begin{equation}\label{35}
  \bar{y}_{k,\nu}(z)=(z+i)^{-\frac{1}{2}(\emph{A}+i\emph{B})}(z-i)^{-\frac{1}{2}(\emph{A}-i\emph{B})}R^{(a,b)}_{\bar{\nu}}(z).
\end{equation}
We may  give the normalization integral as
\begin{equation}\label{41}
  \int^{\infty}_{-\infty}(1+z^{2})^{b-1}e^{a \arctan z}
  R^{(a,b)}_{{\bar{\nu^{'}}}}(z)R^{(a,b)}_{{\bar{\nu}}}(z)dz=\delta^{\bar{\nu^{'}}}_{\bar{\nu}}.
\end{equation}
The solutions of (\ref{33})  gives
\begin{equation}\label{42}
 A^{2}- \omega^{2}=-\bar{\nu}(\bar{\nu}-1+2b)
\end{equation}
and from equations (\ref{42}) and (\ref{sol1}), one can also obtain
\begin{equation}\label{43}
  A^{2}+(m+\frac{1}{2})^{2}-(m+n+\frac{1}{2})^{2}+\bar{\nu}(\bar{\nu}-1+2b)=0.
\end{equation}
According to (\ref{38}) and (\ref{43}), one can express $\ell$ as $\ell_{\nu}$ in terms of quantum numbers. From (\ref{43}), we get
\begin{equation}\label{44}
    A=\nu\pm\sqrt{n+2mn+n^{2}}.
\end{equation}
If we compare (\ref{44}) and (\ref{38}), we have
\begin{eqnarray}
  n^{2}+2mn+n &=& \frac{1}{2} \\
  \nu &=& \frac{\sqrt{2}}{16\ell M \epsilon}\sqrt{\textbf{a}_{1}-1-4\ell^{2}M^{2}} (\textbf{a}_{1}+1+4\ell^{2}M^{2})\label{45}.
\end{eqnarray}
Using (\ref{45}) one can find $\ell$ which is a function of $\nu$.
\section{ Pseudo-Supersymmetry Framework}

Following the fundamental aspects of pseudo-Hermitian quantum mechanics may lead to an outline for understanding spectral aspects of the quantum system. In the light of pseudo-Hermitian generalization of supersymmetric quantum mechanics, one can construct an unknown non-Hermitian Hamiltonian. \\
\textbf{Definition 1:} If $\mathfrak{H}_{\pm}$ are separable Hilbert spaces and an operator $\emph{L}:\mathfrak{H_{+}}\rightarrow \mathfrak{H_{-}}$ is defined and $\eta_{\pm}: \emph{H}_{\pm}\rightarrow \emph{H}_{\pm}$ be  linear  operators which are generally Hermitian and invertible. Then, the pseudo-adjoint of this operator $\emph{L}^{\ddag}: \emph{H}_{-}\rightarrow \emph{H}_{+}$ is equal to $\emph{L}^{\ddag}=\eta^{-1}_{+}\emph{L}^{\dag}\eta_{-}$.\\
\textbf{Definition 2:} Let $\eta_{\pm}=\eta$ which belongs to $\mathfrak{H}_{\pm}=\mathfrak{H}$. $\emph{L}$ is a pseudo-Hermitian operator if
\begin{equation}\label{22}
    \emph{L}^{\dag}=\eta \emph{L} \eta^{-1}.
\end{equation}
\textbf{Definition 3:} $\eta$  is an invertible operator which satisfies $\eta H \eta^{-1}=H^{\dag}$, $H$ is called as pseudo-Hermitian Hamiltonian \cite{mos}. The $\eta$ representation of the Hamiltonian is
\begin{equation}\label{eta}
    H_{\eta}=\eta H \eta^{-1}=H_{\eta}^{\dag}.
\end{equation}
It is noted that the wave function $\psi$ is $\psi=\eta \phi$ and satisfies the wave equation given below
\begin{equation}\label{we}
    H \phi=i\frac{d\phi}{dt},~~~~\hbar=c=1.
\end{equation}
And the scalar product reads
\begin{equation}\label{sp}
    <\xi, \phi>_{\rho}=<\zeta,\psi>=\int \xi^{\dag}\rho \phi~ d\tau,~~~~\rho=\eta^{2}.
\end{equation}
In \cite{3}, the authors constructed unique and self-conjugate Dirac Hamiltonians in gravitational fields using Schwinger gauge. They used an initial Hamiltonian and showed that the system of tetrad vectors remained the same in the $\eta$ representation. Then, $\eta$ is defined by \cite{3},
\begin{equation}\label{eta1}
  \eta=(-\textbf{g})^{\frac{1}{4}}(-\textbf{g}^{00})^{\frac{1}{4}}
\end{equation}
where $\textbf{g}^{00}$ is the time component of the metric tensor and $\textbf{g}=det[g_{\mu\nu}]$. Now following the pseudo-supersymmetric aspects of the system,   $\emph{H}$  is a linear and diagonalizable operator and $Q_{i}$ are linear operators(called as supercharge operators), $i, j=1,2,...,N$, $\tau$ is a grading operator defining the unitary involution
\begin{equation}\label{220}
    \tau^{\dag}\tau=\tau\tau^{\dag}=\tau^{2}=1.
\end{equation}
If $N=1$, we have only one supercharge operator $Q$ which satisfy the algebra \cite{mos}
\begin{equation}\label{221}
  Q^{2}=(Q^{\ddag})^{2}=0,~~~~\{Q,Q^{\ddag}\}=\delta^{j}_{i}\emph{H}.
\end{equation}
According to the ordinary supersymmetric quantum mechanics, two component realization of the system is
\begin{equation}\label{222}
    \tau=\left(
           \begin{array}{cc}
             1 & 0 \\
             0 & -1 \\
           \end{array}
         \right), ~~~~Q=\left(
                          \begin{array}{cc}
                            0 & 0 \\
                            \eta & 0 \\
                          \end{array}
                        \right), ~~~~\emph{H}=\left(
                                                \begin{array}{cc}
                                                  \emph{H}_{+} & 0 \\
                                                  0 & \emph{H}_{-} \\
                                                \end{array}
                                              \right),~~~~\eta=\left(
                                                                 \begin{array}{cc}
                                                                   \eta_{+} & 0 \\
                                                                   0 & \eta_{-} \\
                                                                 \end{array}
                                                               \right)
\end{equation}
where $\emph{H}_{+}=\mathcal{L}_{1}\mathcal{L}_{2}$ and $\emph{H}_{-}=\mathcal{L}_{2} \mathcal{L}_{1}$. In Ref.\cite{roy}, an unknown non-Hermitian  operator $\emph{H}_{+}$ is linked to the adjoint of its pseudo-supersymmetric partner Hamiltonian $\emph{H}_{-}$, i.e.
\begin{equation}\label{223}
    \eta \emph{H}_{+}=\emph{H}^{\dag}_{-}\eta.
\end{equation}
It is noted that $\emph{H}_{+}$ is diagonalizable system admitting a complete biorthonormal system of eigenvectors. If  $\eta_{1}, \eta_{2}$ are the intertwining operators such that
\begin{equation}\label{224}
    \eta_{1}\emph{H}_{+}=\emph{H}_{-}\eta_{1},~~~~\eta_{2}\emph{H}_{-}=\emph{H}^{\dag}_{-}\eta_{2}.
\end{equation}
It can be seen that operator $\emph{H}_{+}$ is $\tilde{\eta}=\eta^{\dag}\eta_{1}$ pseudo-Hermitian and $\emph{H}_{-}$ is $(\eta^{\dag}_{2})^{-1}$ pseudo-Hermitian, $\eta=\eta_{2}\eta_{1}$ \cite{roy} where supercharge operators satisfy  $Q^{\ddag}=\bar{\eta}^{-1}Q^{\dag}\bar{\eta}$ and we give $\bar{\eta}$ and $\tilde{H}$ instead of $\eta$ and $\emph{H}$ in (\ref{222}) as
\begin{equation}\label{225}
 \bar{\eta}=   \left(
      \begin{array}{cc}
        \tilde{\eta} & 0 \\
        0 & (\eta^{\dag}_{2})^{-1} \\
      \end{array}
    \right),~~~~\tilde{H}=\left(
                  \begin{array}{cc}
                   \emph{ H}_{+} & 0 \\
                    0 & \emph{H}_{-}^{\dag} \\
                  \end{array}
                \right).
\end{equation}
If $\emph{H}_{\pm}(\tau)$ are  time dependent pseudo-Hermitian Hamiltonians, then we have
\begin{equation}\label{ham2}
    \emph{H}^{\dag}_{-}=\eta_{2} \emph{H}_{-} \eta_{2}^{-1}+i\eta_{2}^{-1}\frac{\partial \eta_{2}}{\partial \tau},
\end{equation}
\begin{equation}\label{ham20}
   \emph{ H}_{-}=\eta_{1} \emph{H}_{+}\eta^{-1}_{1}+i\eta^{-1}_{1}\frac{\partial \eta_{1}}{\partial \tau}
\end{equation}
and
\begin{equation}\label{ham200}
    \emph{H}_{+}=\eta^{-1}H^{+}_{-}\eta-i\eta^{-1}\frac{\partial \eta}{\partial \tau}.
\end{equation}
The adjoint of (\ref{ham200}) may be put into equation (\ref{ham20}), then one can obtain
\begin{equation}\label{201}
    \eta^{\dag}\eta_{1}\emph{H}_{+}=\emph{H}^{\dag}_{+}\eta^{\dag}\eta_{1}+i(\frac{\partial  \eta^{\dag}}{\partial\tau}\eta_{1}-\frac{\partial \eta_{1}}{\partial \tau}\eta^{\dag}).
\end{equation}
Using (\ref{1}), let us introduce the Hamiltonian as,
\begin{equation}\label{ham1}
  \emph{H}_{-}= i M\ell \gamma^{0}-i\coth \tau-i\frac{\gamma^{0}\gamma^{1}}{\sinh \tau}\partial_{\theta}-i\frac{\gamma^{0}\gamma^{1}}{\sinh \tau}\frac{\cot \theta}{2}-i\frac{\gamma^{0}\gamma^{2}}{\sinh \tau \sin \theta}\partial_{\phi}-i e \ell A_{1}(\tau,\theta)\gamma^{0}\gamma^{1}.
\end{equation}
Thus from (\ref{ham2}) and (\ref{eta1}), we  get
\begin{equation}\label{ham3}
    \emph{H}^{\dag}_{-}=i M\ell \gamma^{0}-i\frac{\gamma^{0}\gamma^{1}}{\sinh \tau}\partial_{\theta}-i\frac{\gamma^{0}\gamma^{2}}{\sinh \tau \sin \theta}\partial_{\phi}-i e \ell A_{1}(\tau,\theta)\gamma^{0}\gamma^{1},
\end{equation}
and we note that
\begin{equation}\label{ham33}
  \eta_{2}= i\ell^{3/2}\sinh \tau \sqrt{\sin \theta},~~~~\eta_{2}^{-1}\frac{\partial \eta_{2}}{\partial \tau}=\coth \tau.
\end{equation}
Let us introduce a form for the metric operator $\eta_1$ as,
  \begin{equation}\label{ph6}
  \eta_1 =\ell^{2}(\sinh\tau)^{a_{1}}(\sqrt{\sin\theta})^{a_{2}}(\sqrt{\sin\phi})^{a_{3}}
\end{equation}
where $a_{1},a_{2},a_{3}$ are real numbers. We may give the unknown Hamiltonian $\emph{H}_{+}$ as
\begin{equation}\label{Harti}
    \emph{H}_{+}=i M\ell \gamma^{0}-i\frac{\gamma^{0}\gamma^{1}}{\sinh \tau}\partial_{\theta}-i\frac{\gamma^{0}\gamma^{2}}{\sinh \tau \sin \theta}\partial_{\phi}-i e \ell A_{1}(\tau,\theta)\gamma^{0}\gamma^{1}+\gamma^{0}\gamma^{1}f[\tau,\theta,\phi]+\gamma^{0}\gamma^{2}g[\tau,\theta,\phi]+U(\tau)
\end{equation}
here $f[\tau,\theta,\phi], g[\tau,\theta,\phi], U(\tau)$ are the unknown functions. Now, both equation (\ref{ham20}) and equation (\ref{ham200}) lead to the following expressions as
\begin{equation}\label{ph7}
  \gamma^{0}\gamma^{1}\left(\frac{1}{2}i\cot\theta\csc h \tau(1+a_{2})+f(\tau,\theta,\phi)\right)=0,
\end{equation}
\begin{equation}\label{ph9}
  \gamma^{0}\gamma^{2}\left(\frac{ia_{3}}{2}\cot\phi \csc\theta \csc h \tau+g(\tau,\theta,\phi) \right)=0,
\end{equation}
\begin{equation}\label{ph90}
   U(\tau)+i(a_{1}+1)\coth\tau=0.
\end{equation}
Choosing specific values such as $a_1=\frac{3}{2}, a_2=1, a_3=1$, we obtain
\begin{equation}\label{Harti}
    \emph{H}_{+}=i M\ell \gamma^{0}-i\frac{\gamma^{0}\gamma^{1}}{\sinh \tau}\partial_{\theta}-i\frac{\gamma^{0}\gamma^{2}}{\sinh \tau \sin \theta}\partial_{\phi}-i e \ell A_{1}(\tau,\theta)\gamma^{0}\gamma^{1}-i\gamma^{0}\gamma^{1}\cot\theta\csc h \tau-i\gamma^{0}\gamma^{2}\cot\phi \csc\theta \csc h \tau-\frac{5i}{2}\coth\tau
\end{equation}
And the metric operator is given by
\begin{equation}\label{etta1}
    \eta_1=\ell^{2}(\sinh\tau)^{3/2}\sqrt{\sin\theta}\sqrt{\sin\phi}=(-\textbf{g}_{1})^{1/4}(-\textbf{g}_{1}^{00})^{1/4}.
\end{equation}
This  also shows that there may be a  metric tensor for the Hamiltonian (\ref{Harti}) which is the partner of $\emph{H}_{-}$. We remind the metric tensor and metric operator for the system  $\emph{H}_{-}$ is given by
\begin{equation}\label{mt1}
    g_{\mu\nu}=\left(
       \begin{array}{ccc}
         \ell^{2} & 0 & 0 \\
         0 & -\ell^{2} \sinh^{2}\tau & 0 \\
         0 & 0 & -\ell^{2} \sinh^{2}\tau\sin^{2}\theta \\
       \end{array}
     \right),~~~~\eta_{2}= i\ell^{3/2}\sinh \tau \sqrt{\sin \theta}.
\end{equation}
Hence, we may obtain $ (\textbf{g}_{1})_{\mu\nu}$ that may be the partner metric tensor of $ g_{\mu\nu}$  as
\begin{equation}\label{mt}
  (\textbf{g}_{1})_{\mu\nu} = \left(
      \begin{array}{cccc}
        -\ell^{2} & 0 & 0 & 0 \\
        0 & \ell^{2}\sinh^{2}\tau & 0 & 0 \\
        0 & 0 & \ell^{2}\sinh^{2}\tau\sin^{2}\theta & 0 \\
        0 &0 & 0 & \ell^{2}\sinh^{2}\tau\sin^{2}\phi \\
      \end{array}
    \right).
\end{equation}
Here, the corresponding metric can be obtained as
\begin{equation}\label{met2}
   ds^{2}=-\ell^{2}d\tau^{2}+\ell^{2}\sinh^{2}\tau d\theta^{2}+\ell^{2}\sinh^{2}\tau\sin^{2}\theta d\phi^{2} +\ell^{2}\sinh^{2}\tau\sin^{2}\phi d\chi^{2}
\end{equation}
and this is the metric of the four-dimensional world in hyper-spherical coordinates.

\newpage
\section{Conclusions}
The Dirac operator is considered in a  three-dimensional gravity and decomposed into  time dependent and angular parts. After performing the separation of variables, the angular Dirac equation is reduced to the  Klein-Gordon- like partner Hamiltonians  possessing shape invariant trigonometric potentials whose real spectrum is given and solutions of the spinor wave-functions are written in terms of the Jacobi polynomials. On the other hand, the time dependent part of the Dirac system  (\ref{30}) and (\ref{31}) have a time dependent potential function whose form is similar to the complex Eckart potential in the literature \cite{Jia}. However,  the function $\csc h^{2}\tau$ has a wrong sign in our case so that  we have used an appropriate mapping  to obtain a soluble hypergeometric differential equation and we have obtained the solutions  in terms of the Romanovski polynomials. At the end of the exact solutions, we have obtained a relation which is a condition between the quantum numbers of the angular and time dependent parts and the system's parameters. The radius of the universe is given by $\ell$ and we have seen that this radius depends on $\nu$ quantum number. Moreover, we have shown that a new $(3+1)$ dimensional  Hamiltonian $\emph{H}_{+}$ can be obtained  by means of the pseudo-supersymmetric procedure which means one can expand $(2+1)$ Dirac system to $(3+1)$ dimensions. Because two partner Hamiltonians are pseudo-Hermitian,  we have found the metric operator $\eta_{1}$ that links $\emph{H}_{\pm}$. We have also seen that another metric tensor for the partner Hamiltonian $\emph{H}_{+}$ may be obtained using pseudo-supersymmetry. In our case, the metric operators $\eta_{1}$ and $\eta_{2}$ are not differential operators. We have used an ansatze for the $\eta_{1}$ metric operator, however, if there is a symmetry in curved Dirac systems which gives metric operator without giving an ansatze can be searched in the future works.

\end{document}